\documentclass[10pt]{article}
\usepackage{amssymb,euscript,bbold}
\usepackage{amsfonts}
\usepackage{amssymb}
\usepackage{mathrsfs}
\usepackage{graphicx}
\usepackage{slashed}
\usepackage{amsthm}
\usepackage{amsmath}
\UseRawInputEncoding
\newcommand{\be}{\begin{equation}}
\newcommand{\ee}{\end{equation}}
\newcommand{\ba}{\begin{eqnarray}}
\newcommand{\ea}{\end{eqnarray}}

\usepackage{xcolor}
\usepackage{bbold} 
\usepackage{hyperref}

\begin{document}
\input{epsf}

\begin{flushright}
KCL-2024-039
\end{flushright}
\begin{flushright}
\end{flushright}
\begin{center}
\Large{Confinement in Five Dimensions}\\
\bigskip
\large{B.S. Acharya}\\
\smallskip\normalsize{\it
Abdus Salam International Centre for Theoretical Physics, Strada Costiera 11, 34151, Trieste, Italy}\\

and

{\it Department of Physics, Kings College London, London, WC2R 2LS, UK}\\
\end{center}

\bigskip
\begin{center}
{\bf {\sc Abstract}}

Five dimensional super conformal field theories can be studied using their geometric realisation as a limit of $M$-theory on a metrically conical Calabi-Yau threefold. We utilise this framework to investigate the phases of such theories that arise by varying the couplings away from the conformal point. We demonstrate that many 5d SCFTs, including strongly coupled gauge theories, have couplings giving rise to massive, confining vacua with confining strings and corresponding unbroken 1-form symmetries. The simplest examples arise by considering the parameter space of {\it complete} Ricci flat metrics on discrete quotients of the standard conifold singularity. Varying other couplings produces coupled 5d SCFTs interacting via massive BPS instanton particle states.

\end{center}

\newpage

\section{Introduction.}

A remarkable fact about string/$M$-theory concerns the physics of spacetime singularities. If space develops certain, rather special kinds of singularity, light, interacting degrees of freedom can arise, typically described by an interacting conformal field theory. One might therefore like to understand more generally which singularities in spacetime make sense physically and have this kind of interpretation, perhaps eventually providing an interpretation of cosmological singularities, though we are rather far from such an understanding today.

The existence of interacting conformal field theories in five and six dimensions is also a remarkable fact which was first discovered in the string/$M$-theory framework. In particular, five dimensional superconformal theories (SCFTs) were discovered by Seiberg \cite{Seiberg:1996bd}. These were then interpreted in \cite{Morrison:1996xf, Intriligator:1997pq, Douglas:1996xp} as arising from $M$-theory on Calabi-Yau threefolds with certain special kinds of singularity. Since then, this geometric framework has been one of the main tools for studying such theories.

In this paper we will study the geometric $M$-theory description of the conformal theories which arise at strong coupling in five dimensional super Yang-Mills theory with ADE gauge group (and zero Chern-Simons term). Said differently, we will study geometrically the conformal field theories which admit deformations whose infrared limit are five dimensional super Yang-Mills theories.
The singular Calabi-Yau spaces corresponding to the conformal fixed points turn out to be discrete quotients of the standard conifold singularity with its standard Ricci flat cone metric \cite{Candelas:1989js}. This enables a simple analysis of the physics as we vary the various coupling constants. As we will show, {\it in addition to the Yang-Mills coupling constant, there are two additional couplings whose roles are privileged in the geometric description}. 
In particular, there is a one parameter family of complete, asymptotically conical Calabi-Yau metrics on a smooth 6-manifold which desingularises the space and hence this parameter deforms the conformal field theory. Along this branch, an analysis of the spacetime metric reveals that there are {\it no $L^2$-normalisable zero modes and hence the resulting theory is massive}. However, the 6-manifold has a non-trivial fundamental group and correspondingly the five dimensional theory has stable strings whose charges are given by the centre of the original gauge group, which shows that {\it the theory has an unbroken 1-form symmetry and exhibits confinement at low energies}. The strings are $M$2-branes which wrap incontractible loops in the extra dimensions.

The results of this paper are the five dimensional analogues of the four dimensional theories which arise by studying $M$-theory on certain asymptotically conical spaces of $G_2$-holonomy. There one was able to study super Yang-Mills theory geometrically and prove that the theory has a mass gap and exhibits confinement \cite{bsa2,Atiyah:2000z,Atiyah:2001qf,Acharya:2001hq,Acharya:2001dz}.

Having studied these examples, for completeness we also study the theories obtained from general quotients of the conifold. These also deform to produce confining vacua in five dimensions as well as new five dimensional theories which consist of pairs of interacting SCFTs which are coupled together non-conformally and contain massive BPS instanton particles whose mass goes to zero at the SCFT point. The results suggest that the existence of massive, confining deformations of five dimensional conformal theories might be a fairly common phenomenon.


The main results of this paper are summarised in the figure. In the geometric approach to 5d SCFTs, strong coupling fixed points are characterised by special codimension six {\it conical} singularities in spacetime. For the singularities corresponding to the conformal theories studied in this paper, we will see that they can be reached in three different ways via degenerations of three topologically distinct, smooth, complete spacetimes. Physically this implies that there are three different low energy theories that arise by deforming the same conformal theory, indicated by the three lines in the figure. Since the geometric analysis is only really valid suitably far out along the three lines, we can't really say what happens in the centre of the figure, apart from saying that the CFT point exists there.

In the figure, the conformal field theory is indicated by the cyan disk at the centre. There are three distinguished couplings, $a_1, a_2$ and $\mu$ with $(a_1 , a_2, \mu ) = (0,0,0)$ being the conformal point. Switching on these parameters individually we obtain three different theories at low energies. The red line is the weakly coupled gauge theory. Far along the blue line we obtain a massive, confining theory. 
Finally, along the green line the system consists of a pair of conformal theories which are non-conformally coupled via massive BPS instanton particles.

\begin{figure}
    \centering
    \includegraphics[width=0.5\linewidth]{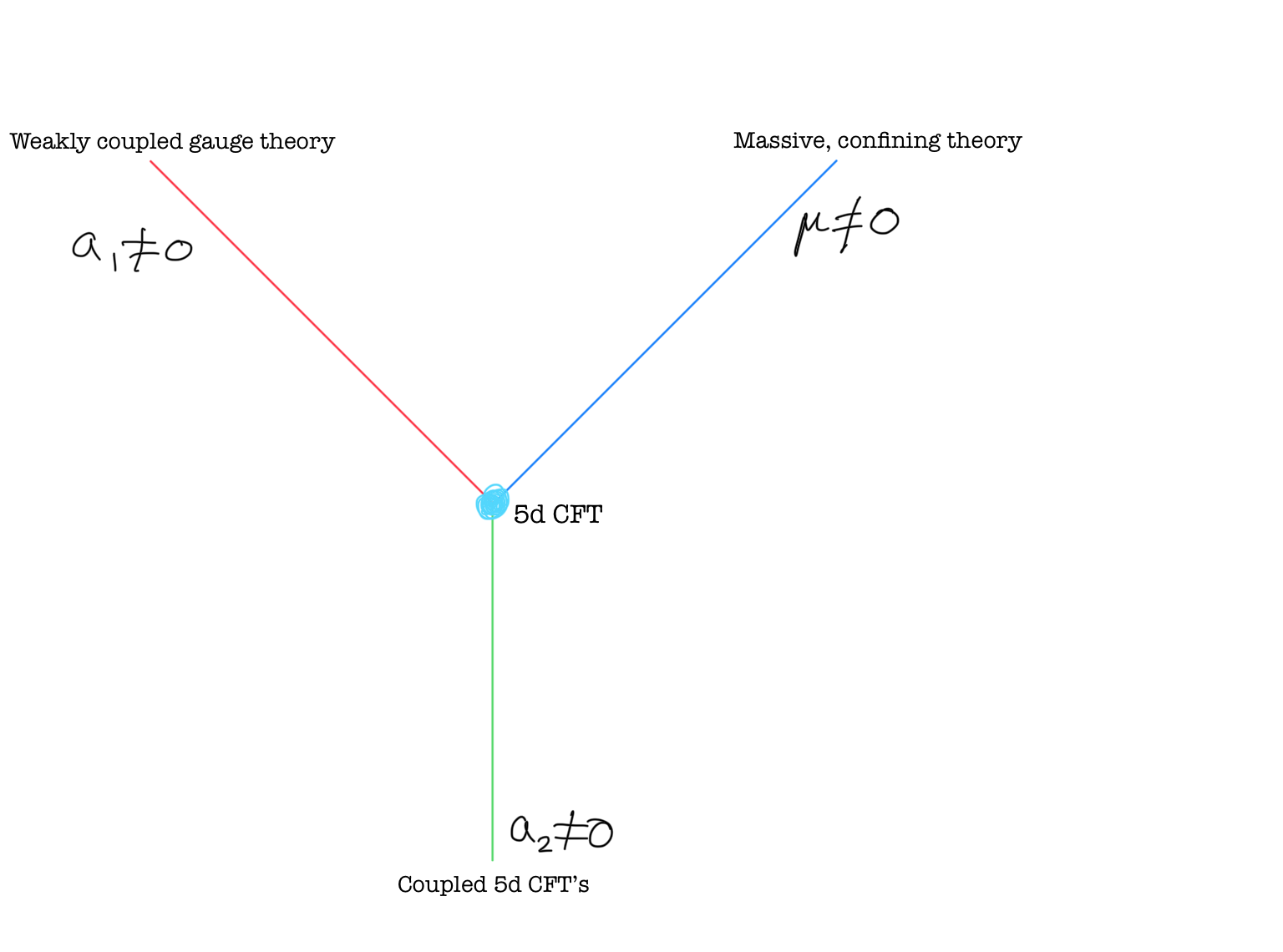}
    \caption{Three geometric couplings of the 5d CFT and their weak coupling interpretations. }
    \label{fig:enter-label}
\end{figure}

The rest of the paper begins with a review of the geometry of Calabi-Yau conifolds, their symmetries and their parameters.  We then describe in detail the complete asymptotically conical Calabi-Yau spaces which correspond to weakly coupled super Yang-Mills theory and describe the geometry of the strong coupling limit. Next, we analyse the parameter space emanating from the strong coupling limit and show that there are two additional parameters beyond the Yang-Mills coupling that deform the theory in non-trivial ways whose interpretation is given. In section four one of these deformations is shown to produce massive confining theories whilst in section five the other deformation is shown to produce the coupled conformal theories. We close the paper with a preliminary analysis of the more general theories that arise in this framework.

\section{5d Superconformal Theories and Calabi-Yau Cones}

We will consider $M$-theory spacetimes of the form $M = Z \times \mathbb{R}^5$ with, $Z$ a 6-manifold with a background metric $g(M) = g(Z)+\eta$, where $g(Z)$ is a Calabi-Yau metric on $Z$ and $\eta$ is the flat Minkowski metric. When $Z$ is compact, the low energy effective theory iss a five-dimensional supergravity theory on $\mathbb{R}^5$ with eight supercharges. This theory has a moduli space of vacua, closely related to the moduli space of Calabi-Yau metrics on $Z$. At special points in the moduli space, $Z$ may develop very particular kinds of singularities and additional light degrees of freedom arise there. We are interested in the conical singularities that $Z$ may develop since these are expected to correspond to 5d superconformal field theories. Near such a singularity, $Z$ looks like $M \cong \mathbb{R}^+ \times \Sigma$ , with $\Sigma$ a compact 5-manifold. The metric on $M$ looks conical around this point: $ds_0^2 = dr^2 + r^2 g(\Sigma)$ and it has holonomy group $SU(3)$. 

In order to {\it completely} decouple gravity from the degrees of freedom localised at the conical singularity, one apparently requires a {\it complete} asymptotically conical Calabi-Yau metric on a non-compact 6-manifold, $M$, which looks like $ds_0^2 = dr^2 + r^2 ds(\Sigma)^2$ as $r$ becomes very large. Additionally, physically, $M$ (and its five-dimensional cross-section $\Sigma$) are allowed to have orbifold singularities and the metric is required to be complete and Calabi-Yau in the orbifold sense. We are thus interested complete asymptotically conical Calabi-Yau orbifolds. This is the precise definition of the geometric framework we will study. 
Note that asymptotically conical Calabi-Yau manifolds have recently been classified \cite{Conlon:2022nug}. Much of the existing physics literature on the nature of the singularities of Calabi-Yau threefolds corresponding to five dimensional theories is rather topological and algebraic in flavour with very little reference to the Calabi-Yau metric. Instead we will insist on the completeness and asymptotically conical criteria which correspond to decoupling gravity and conformality respectively. These properties of the metric will actually lead to some of the most physically significant results in what follows.

\section{5d Super Yang-Mills and Hyperconifolds}

In Yang-Mills theories in five dimensions, $1/g^2_{YM}$ has mass dimension one. Consequently, any perturbative process is suppressed by the mass scale $1/g^2_{YM}$ and becomes less (or more) probable as the momenta/energies of external particles are decreased (or increased). The theory is thus weakly coupled at low energies. One remarkable fact is that supersymmetric Yang-Mills theory at weak coupling arises by deforming a strongly coupled interacting super conformal field theory. We now describe the geometric realisation of these theories in the $M$-theory framework. In order to do that we will need to review known facts about conifolds, their parameters and their symmetries.

\subsection{The Conifold, its Desingularisations and Calabi-Yau metrics}

We will require several facts about the conifold and its resolutions and smoothings. Most of the material in this subsection is in the original paper of Candelas-de la Ossa \cite{Candelas:1989js}.
The conifold, ${\cal C}_0$,  is a six-dimensional space with an isolated singular point which admits a conical Calabi-Yau metric. One can describe ${\cal C}_0$ topologically as $\mathbb{R}^+ \times S^2 \times S^3$ as well as algebraically as a quadric in $\mathbb{C}^4$:
\begin{equation}
    P_0 : z_1^2 + z_2^2 + z_3^2 + z_4^2 = 0
\end{equation}

The conical Calabi-Yau metric on ${\cal C}_0$  is given by
\begin{equation}
    g({\cal C}_0) = dr^2 + r^2 \;g_H(S^2 \times S^3)
\end{equation}
where  $g_H(S^2 \times S^3)$ is the homogeneous Einstein metric on $(SU(2) \times SU(2))/U(1)$, with the $U(1)$ embedded symmetrically in each $SU(2)$, both regarded as degree one line bundles over $SU(2)/U(1)$. $r$ is a coordinate on $\mathbb{R}^+$ and the space has a conical singularity at $r=0$. Since a certain $\mathbb{Z}_2$ subgroup acts trivially, the metric has an $SO(4) = (SU(2) \times SU(2))/\mathbb{Z}_2$ group of symmetries.

${\cal C}_0$ has three distinct desingularisations to smooth 6-manifolds, ${\cal C}_{a_1}$, ${\cal C}_{a_2}$  and ${\cal{C}}_\mu$, labelled by parameters $(a_1, a_2, \mu) \in (\mathbb{R}^+, \mathbb{R}^+, \mathbb{C})$. These are each obtained by removing a neighbourhood of the singular point in ${\cal C}_0$ and replacing it with a suitable compact set whose boundary matches that of ${\cal C}_0 \setminus \{0\}$. Hence the boundaries of these three manifolds are all topologically $S^2 \times S^3$. The parameters $(a_1, a_2, \mu) \in (\mathbb{R}^+, \mathbb{R}^+, \mathbb{C})$ have magnitudes which fix the size of the compact region and are free parameters. We will need to describe these three manifolds in more detail, since these three parameters will eventually be the coupling constants of the 5d theories we discuss.

\bigskip

\begin{itemize}
    \item {\underline{$\mu \neq 0$:}}    ${\cal{C}}_\mu$  is simply obtained by deforming the algebraic equation to
\end{itemize}

\begin{equation}\label{quadric}
    P_{\mu}: z_1^2 + z_2^2 + z_3^2 + z_4^2 = \mu
\end{equation}

This explicitly shows that ${\cal{C}}_\mu = T^*(S^3)$, the cotangent bundle of the sphere and that the manifold contains a compact $S^3$ of radius $|\mu|^{1/2}$ at the centre.

\bigskip

\begin{itemize}
    \item { \underline{$a_1 \neq 0$:} To describe ${\cal C}_{a_1}$}, we rewrite the quadric as $\det(W)=0$, with $W$ the matrix
\end{itemize}

\begin{equation}
W = \begin{pmatrix}
X&U\\V&Y
\end{pmatrix} = {1 \over \sqrt{2}}\begin{pmatrix}
z_1 + iz_2 &-z_3 + i z_4\\ z_3 + i z_4 &z_1 - i z_2
\end{pmatrix}
\end{equation}
and one replaces it with the simultaneous solutions of $\det(W)=0$ and

\begin{equation}
W = \begin{pmatrix}
X&U\\V&Y
\end{pmatrix} \begin{pmatrix}
    \lambda_1 \\ \lambda_2
\end{pmatrix}
\end{equation}
where $(\lambda_1, \lambda_2)$ are coordinates on $\mathbb{CP}^1$. This determines a point in $\mathbb{CP}^1$ for every non-zero point on ${\cal C}_{0}$, whilst the origin is replaced by the entire $\mathbb{CP}^1$.  ${\cal C}_{a_1}$ is in fact a rank two complex vector bundle over $\mathbb{CP}^1$ which is the sum of two degree minus one line bundles, ${\cal C}_{a_1} = ({\cal O}(-1)\oplus {\cal O}(-1))|_{\mathbb{CP}^1}$. The real parameter $a_1$ gives the size of the $\mathbb{CP}^1$ at the origin and will be important when we discuss the physics.

Since it is a vector bundle over $\mathbb{CP}^1$, ${\cal C}_{a_1}$ can be covered by two coordinate patches, $\lambda_1 \neq 0$, $\lambda_2 \neq 0$. Explicitly, these are
\begin{equation}
    {\cal C}_{a_1} = U_{\lambda_1 \neq 0} \cup U_{\lambda_2 \neq 0} = \{ X,V, U/X = Y/V \} \cup \{ Y,U, X/U = V/Y \} 
\end{equation}

\bigskip

\begin{itemize}
    \item { \underline{$a_2 \neq 0$}: Similarly, to describe ${\cal C}_{a_2}$},
\end{itemize} we again consider $\det(W)=0$ but blow up a different $\mathbb{CP}^1$:

\begin{equation}
W = \begin{pmatrix}
    \lambda_1 & \lambda_2
\end{pmatrix}\begin{pmatrix}
X&U\\V&Y
\end{pmatrix} 
\end{equation}
where $(\lambda_1, \lambda_2)$ are coordinates on another $\mathbb{CP}^1$. This is a distinct resolution but with the same topology ${\cal C}_{a_2} = ({\cal O}(-1)\oplus {\cal O}(-1))|_{\mathbb{CP}^1}$. Coordinate charts can be given by

\begin{equation}
    {\cal C}_{a_2} = U_{\lambda_1 \neq 0} \cup U_{\lambda_2 \neq 0} = \{ X,U, V/X = Y/U \} \cup \{ Y,V, X/V = U/Y \} 
\end{equation}

\bigskip

{\subsection{Asymptotically Conical Calabi-Yau metrics:}}

Candelas and de la Ossa \cite{Candelas:1989js} showed that all three manifolds admit smooth, complete Calabi-Yau metrics, $g_{a_1}, g_{a_2}$ and $g_{\mu}$ all of which asymptote to the conical metric at infinity. These metrics also have the full $SO(4)$ symmetry of the conical model. Furthermore, the parameters $a_1, a_2,\mu$ control the sizes of the respective spheres at the origin, which turn out to be volume minimising supersymmetric (i.e. calibrated) cycles.

Finally we can come to the physical theories of interest in this paper.  They all arise by considering discrete quotients of these three spaces and the conifold by subgroups of the $SO(4)$ isometry group. Such quotients have previously been considered by Gubser et al. \cite{Gubser:1998ia} and Davies \cite{Davies:2011is,Davies:2013pna} and the results of this paper provide a physical interpretation of some of the results and ideas of those papers.

\bigskip

\subsection{Super Yang-Mills from $({{\cal O}(-1)\oplus {\cal O}(-1) \over {\Gamma_\mathbb{ADE}}})|_{{\mathbb CP}^1}$}

As a 6-manifold, ${\cal C}_{a_1} = \mathbb{R}^4 \times S^2$ and the quotients we will initially consider replace each $\mathbb{R}^4$ fibre by the orbifold ${\mathbb{R}^4}/{\Gamma_{\mathbb{ADE}}}$, where $\small{\Gamma}_\mathbb{ADE}$ is a finite subgroup of $SU(2)$ which fixes the $0 \times S^2$ at the origin in  $\mathbb{R}^4 \times S^2$.  Since the Candelas-de la Ossa confold metric $g_{a_1}$ is $\small{\Gamma}_\mathbb{ADE}$-invariant,  ${\cal C}_{a_1}/\small{\Gamma}_\mathbb{ADE} = ({{\cal O}(-1)\oplus {\cal O}(-1) \over {\Gamma_\mathbb{ADE}}})|_{{\mathbb CP}^1} = {\mathbb{R}^4}/\small{{\Gamma}}_\mathbb{ADE} \times S^2$  {\it  inherits a complete asymptotically conical Calabi-Yau orbifold metric, asymptotic to a metric cone over} $(S^2 \times S^3)/\Gamma_\mathbb{ADE}$.

We can see this more explicitly by calculating the action of $\small{\Gamma}_\mathbb{ADE}$ on  ${\cal C}_{a_1}$ using the previous subsection. The original coordinates $z_i$ transform in the fundamental representation of $SO(4)$. The $\small{\Gamma}_\mathbb{ADE}$ discrete subgroups of interest are defined as the finite subgroups of $SU(2)_L \subset SO(4) \equiv (SU(2)_L \times SU(2)_R)/\mathbb{Z}_2)$.  $SU(2)_L$ acts on the left on the pair $(X,V)$ in the fundamental representation. Similarly, it acts on $(Y,-U)$ in the conjugate representation. Because of this, $SU(2)_L$ acts trivially on the $\mathbb{CP}^1$ coordinates $X/U$ and $Y/V$. To show this we parametrise $SU(2)_L$ matrices $g_L$ as
\begin{equation}
g_L = \begin{pmatrix}
a&b\\-b^* &a^*
\end{pmatrix} 
\end{equation}

with $aa^* + bb^* =1$. Then 

\begin{equation}
X/U \longrightarrow (aX+bV)/(bY+aU) = (aX+bXY/U)/(bY+aU) = X/U
\end{equation}

This proves that  ${\cal C}_{a_1}/\small{\Gamma}_\mathbb{ADE} = ({{\cal O}(-1)\oplus {\cal O}(-1) \over \small{\Gamma}_\mathbb{ADE}})|_{{\mathbb CP}^1} = {\mathbb R}^4 /\small{\Gamma}_\mathbb{ADE} \times S^2$. For simplicity, consider the examples $\Gamma_{\mathbb{A}} = \mathbb{Z}_N$. The action on the coordinates is then given by 
\begin{equation}
\mathbb{Z}_N: (X,Y,U,V) \longrightarrow (\bar{\alpha}X,\alpha Y, \bar{\alpha} U, {\alpha}V)
\end{equation}
where $\alpha^N =1$ is a primitive $N$-th root of unity.

Hence, under the $\mathbb{Z}_N$ action the local coordinates in the two patches transform with weights which are $(-1,1,0)$ and $(1,-1,0)$ respectively, which shows that the $\mathbb{CP}^1$ at the origin i.e. the zero-section is fixed by $\mathbb{Z}_N$. The discussion is similar for all $\Gamma_\mathbb{ADE} \subset SU(2)_L$.

At low energies, $M$-theory on ${\cal C}_{a_1} /\Gamma_\mathbb{ADE} = ({{\cal O}(-1)\oplus {\cal O}(-1) \over {\Gamma_\mathbb{ADE}}})|_{{\mathbb CP}^1}$ is described by five dimensional super Yang-Mills theory with ADE gauge group: the ADE singularity is along a copy of $\mathbb{CP}^1 \times \mathbb{R}^5$ and supports a 7d ADE super Yang-Mills theory. At low energies, integrating over the $\mathbb{CP}^1$ we obtain the fields of the pure 5d super Yang-Mills theory. The gauge coupling of the 5d theory is given, in 11d units, by ${1 \over g^2_{YM}} \propto 4\pi a_1^2$, the volume of the minimal $\mathbb{CP}^1$ in the Candelas-de la Ossa metric. Thus, the theory is weakly coupled at large $a_1$.

\bigskip

{\it  Massive BPS particles and monopole strings. }

\bigskip

The super Yang-Mills theory has a Coulomb branch of vacua along which the adjoint scalars take non-zero vacuum expectation values and break the gauge symmetry to the maximal torus $U(1)^{rk(ADE)}$. This produces massive vector multiplets containing $W$-bosons. These are in fact BPS states. In $M$-theory this can be seen by realising that along the Coulomb branch the $\mathbb{CP}^1$ of ADE singularities gets resolved by a configuration of rk(ADE) copies of $S^2 \times \mathbb{CP}^1$, which intersect according to the ADE Dynkin diagram. The rk(ADE) independent Coulomb vevs, $v_i$ are the differences of the volumes of these 2-spheres\footnote{For small $v_i$ there exists a smooth, complete asymptotically conical Calabi-Yau metric and the 2-sphere volumes are minimal wrt to this metric.} In particular, each of these 2-spheres minimises volume within its homology class. The $W$-boson particles are identified as $M2$-branes which wrap these various spheres and are BPS precisely because the spheres are volume minimising within homology - they give the smallest mass particles for a given charge. $m_W = q|v|$ is the schematic formula for the mass-to-charge ratio.

Similarly, $M2$-branes which wrap the original $\mathbb{CP}^1$ at the origin are also BPS particles. These are the so-called instantonic particles of the 5d theory since they are charged under the $U(1)$ global symmetry associated to the conserved current, $j_I = \star Tr (F \wedge F)$ \cite{Seiberg:1996bd} and give rise to 4d instantons upon compactification to four dimensions. The instanton particle masses at generic points on the Coulomb branch is  $m_I = 1/g^2_{YM}$ , though they can fractionate at the origin. See \cite{Lambert:2014jna, Tachikawa:2015mha} for other aspects of these instantons. In $M$-theory, along the Coulomb branch where the Calabi-Yau threefold is smooth, the current $j_I$ arises from the $C$-field equation of motion integrated over the Calabi-Yau.

Finally, there are BPS monopole strings in the spectrum. These arise in $M$-theory as $M$5-branes wrapping the 4-manifolds of the form $S^2 \times \mathbb{CP}^1$. The monopole tensions are given by an exact formula which is schematically $T_m = |v|/g^2_{YM}$.

The conformal theory of interest arises when all of these BPS particles and strings become massless, namely when all the $v_i$ and $1/g^2_{YM}$ go to zero. The fact that electric, magnetic and instanton particle states are becoming massless simultaneous seems to be a ubiquitous signal for the appearance of a conformal fixed point at strong coupling. In this limit, ${\cal C}_{a_1} /\small{\Gamma}_\mathbb{ADE}$ formally degenerates to the hyperconifold ${\cal C}_0/\small{\Gamma}_\mathbb{ADE}$ with the conifold metric $g_0$. The key statement is that $M$-theory on the hyperconifold ${\cal C}_0/\small{\Gamma}_\mathbb{ADE}$ with the conifold metric $g_0$ gives rise to a 5d SCFT which is the 5d ADE super Yang-Mills theory at strong coupling.

From this point of view, the weakly coupled super Yang-Mills theory is obtained from the 5d SCFT by switching on the coupling $1/g^2_{YM}$ . Geometrically it corresponds to desingularising ${\cal C}_0/\small{\Gamma}_\mathbb{ADE}$ to ${\cal C}_{a_1}/\small{\Gamma}_\mathbb{ADE}$ with the Candelas-de la Ossa metric, $g_{a_1}$.  This makes sense because $g_{a_1}$ is both complete and asymptotic to $g_0$ at infinity. However, as we have reviewed above, there are two  other complete metrics $g_{a_2}$ and $g_\mu$ asymptotic to $g_0$ at infinity and these also admit the $\Gamma_\mathbb{ADE}$ symmetry. Hence the parameters $a_2$ and $\mu$ also correspond to couplings in the 5d theory and one can ask about the physics of these deformations of the 5d conformal theories.
 
\section{Confinement in Five Dimensions}

 We now consider the theory for $\mu \neq 0$, which leads to the main results of this paper.

In this case we obtain $M$-theory on a {\it smooth} 6-manifold which is ${{\cal{C}}_\mu}/\small{\Gamma}_\mathbb{ADE} = T^*({S^3}/\small{\Gamma}_\mathbb{ADE})$, again with a complete, asymptotically conical Calabi-Yau metric, $g_{\mu}$.  This is clear from the description of ${\cal C}_{\mu}$ as a quadric \ref{quadric}:  $\Gamma_\mathbb{ADE}$ acts freely on the sphere when $\mu \neq 0$. What is the physics of $M$-theory on ${{\cal{C}}_\mu}/\small{\Gamma}_\mathbb{ADE}$ with this metric? The background spacetime is completely smooth so is amenable, at suitably large $\mu$, to a field theory analysis.

In analysing the low energy spectrum around a fixed, smooth Calabi-Yau background, massless states arise from harmonic 2-forms and 3-forms. However, since ${{\cal{C}}_\mu}/\small{\Gamma}_\mathbb{ADE}$ is non-compact, we are actually interested in the $L^2$-normalisable harmonic forms with respect  to the Candelas-de la Ossa metric $g_{\mu}$ , since it is these which give rise to massless fields in five dimensions with finite kinetic terms. In actual fact, {\it  there are none of these}, as can be seen from the analysis of \cite{Lockhart} and \cite{Hausel:2002xg} which analyses the $L^2$-cohomology of asymptotically conical spaces in general.

{{\bf Theorem} (\cite{Lockhart}, \cite{Hausel:2002xg}):}
{\it If $(M,g)$ is an asymptotically conical Riemannian n-manifold, the $L^2$-cohomology of $(M,g)$ is given by}:
\ba
L^2 H^{k}(M) \;\;\;\;\; & \cong & \;\;\;\;\; H^{k}(M,\partial M), \;\; k < n/2\\
L^2 H^{n/2}(M)  \;\;\;\; & \cong & \;\;\;\;\; Im(H^{n/2} (M,\partial M) \rightarrow H^{n/2} (M)) \\
L^2 H^{k}(M) \;\;\;\; & \cong & \;\;\;\;\;\; H^{k}(M), \;\; k > n/2  
\ea

In our case, $M$ is a quotient of $T^*(S^3)$ and $\partial M $ a quotient of $S^2 \times S^3$ so its real cohomology comes from that of  $T^*(S^3)$. To compute the cohomology groups above one can use the long exact cohomology sequence for the pair $(M, \partial M)$:

\begin{equation}
    \dots. H^p(M, \partial M ) \rightarrow H^p (M) \rightarrow H^p (\partial M) \rightarrow H^{p+1} (M, \partial M ) \rightarrow \dots
\end{equation}

In our example, $(M, \partial M) = (T^*(S^3), S^2 \times S^3 )$ and explicitly working out the exact sequences we learn that the entire $L^2$-cohomology is empty. For instance, to compute the 3rd $L^2$-cohomology we consider the piece of the above sequence with $p=3$.
Since, in the case at hand, the map from $H^3(T^*(S^3))$ to $H^3(S^2 \times S^3)$ is actually an isomorphism and the kernel is trivial, which shows that the image of  $H^3(M, \partial M )$
in $H^3(M)$ is trivial in this case and hence there are no $L^2$-normalisable harmonic 3-forms on $(C_{\mu}, g_\mu)$.

Intuitively, the Poincare dual of the compact 3-cycle, $S^3$ in $M$ is a compactly supported harmonic 3-form, $\alpha$. The image of $\alpha$ in the $L^2$-cohomology is actually proportional to the self-intersection number of the zero-section $S^3$ in ${\cal C}_\mu$. However, this is just the Euler number of $S^3$, which is zero. In fact, explicit calculation shows that $\alpha$ has a 
norm which diverges logarithmically \cite{Cvetic:2000db}.

Physically this means that the theory for non-zero $\mu$ has a mass gap. However, the theory is still interesting as it has an unbroken 1-form symmetry (in all cases except $\Gamma_{E_8}$). This is due to the fact that $M$2-branes wrapped on non-trivial 1-cycles appear as stable strings in five dimensions and these have charges classified by the first homology group. This group, as was first shown in \cite{Acharya:2001hq}, is $H_1({{\cal{C}}_\mu}/\small{\Gamma}_\mathbb{ADE}) \cong Z(ADE)$ the centre of the compact, simply connected ADE group. Therefore these five dimensional theories are confining at low energies. The $E_8$ theory does not confine as $E_8$ has trivial centre. 

Notice that all of the BPS particles and monopole strings from the Yang-Mills theory have disappeared when $\mu \neq 0$: there are no compact, incontractible even-dimensional cycles in $C_{\mu}/\Gamma_\mathbb{ADE}$ .
This is intuitively consistent with the confinement of the original electric degrees of freedom. Furthermore, the instanton symmetry has been explicitly broken suggesting that the deformation to the confining phase involves the instanton operators explicitly and is reminiscent of confinement in three dimensional QED where monopole operators induce the confining phase \cite{Polyakov:1975rs}. In \cite{Cordova:2016xhm}, the authors have analysed all supersymmetric operator deformations of these 5d SCFTs and it would be very interesting to connect the results here with those as well as with the results on the Higgs branch of the SCFTs in \cite{Aharony:1997bh, Cremonesi:2015lsa}. 

There are however two classes of BPS objects in this theory: instantons and membranes. The instantons arise from Euclidean $M$2-branes wrapping the supersymmetric $S^3/\small{\Gamma_\mathbb{ADE}}$ at the centre. These will certainly give rise to effects of order $\exp(-c|\mu|^3)$ and hence might also be consistent with a small dynamically generated scale. For large $\mu$, which is where our analysis is valid, we can neglect the effects of these instantons. Their presence, however, suggests that the formal $\mu \rightarrow 0$ limit could have an interesting structure since in principle the geometry could be deformed from the classical one by the instanton effects.

BPS  membranes arise from $M$5-branes which wrap the minimal volume $S^3/{{\Gamma}_\mathbb{ADE}}$ in the centre of the manifold.
The world-volume theory of these branes is a three dimensional Yang-Mills Chern-Simons theory with four supercharges. In fact it is the twice supersymmetric version of the theory first discussed in \cite{Acharya:2001dz} as governing the dynamics of domain walls in four dimensional pure supersymmetric Yang-Mills theories. These theories have a non-zero Witten index and, hence, there are non-trivial bound states formed by these membranes. This suggests that there is a non-trivial 2-form generalised symmetry in these theories and aspects of this could be analysed using the techniques of 
\cite{Apruzzi:2021nmk, Cvetic:2022imb, DelZotto:2022fnw}.

\section{$M$-theory on ${\cal C}_{a_2}/\small{\Gamma}_\mathbb{ADE}$: Coupled 5d SCFTs}

Next we consider the theory obtained by deforming the SCFT by $a_2$. ${\cal C}_{a_2}$ can also be covered by two coordinate charts:

\begin{equation}
    {\cal C}_{a_2} = U_{\lambda_1 \neq 0} \cup U_{\lambda_2 \neq 0} = \{ X,U, V/X = Y/U \} \cup \{ Y,V, X/V = U/Y \} 
\end{equation}

Hence, under the $\mathbb{Z}_N$ symmetry the two coordinate charts have weights $(-1,-1,2)$ and $(1,1,-2)$. This is very different to the action on ${\cal C}_{a_1}$. Unlike in section three, here the symmetry acts non-trivially on the $\mathbb{CP}^1$ zero section. Therefore, if $N$ is prime, the singularities of ${\cal C}_{a_2}/\mathbb{Z}_N$ consist of two isolated fixed points, located at the north and south poles of the $\mathbb{CP}^1$ and each modelled on
the singularity at the origin in $\mathbb{C}^3 /\mathbb{Z}_N$ given in coordinates as
\begin{equation}
    (z_1 , z_2, z_3) \rightarrow (\alpha z_1 , \alpha z_2 , \alpha^{-2} z_3 )
\end{equation}
Now, $M$-theory on $\mathbb{C}^3 /\mathbb{Z}_N$ itself produces a 5d SCFT located at the origin e.g. $N=3$ is the mysterious Seiberg $E_0$ theory \cite{Seiberg:1996bd}. Therefore, in $M$-theory on  ${\cal C}_{a_2}/\mathbb{Z}_N$ one has a pair of $\mathbb{C}^3 /\mathbb{Z}_N$ SCFTs which are coupled via the non-zero size $\mathbb{CP}^1$ of radius $a_2$. Notice also that if $N$ is even, then the order two elements of $\mathbb{Z}_N$ fix the $\mathbb{CP}^1$ and the low energy theory also includes $SU(2)$ gauge fields with $a_2 \sim 1/g_{YM}$. This demonstrates that $a_2$ is  a coupling in these models. The theory also contains instanton particles which wrap (the $\mathbb{Z}_N$-quotient of) the $\mathbb{CP}^1$ and couple to the two 5d theories.  Since the two coupled SCFTs are isomorphic there is a one-to-one correspondence between their operators and the geometry strongly suggests that the coupling between them when $a_2 \neq 0$ respects the symmetry between them i.e. only diagonal couplings are present.

Notice that in the $N=2$ case, the $\mathbb{Z}_2$ fixes the $\mathbb{CP}^1$ and the theory is isomorphic to the $E_1$ theory.

\section{More general quotients and theories}

Here we briefly consider the theories obtained by considering more general quotients of the conifold, beyond $\Gamma_\mathbb{ADE}$. Therefore we wish to consider generic finite subgroups $\Gamma \subset SO(4)$. A similar analysis was given in \cite{Acharya:2023bth} for four dimensional $N=1$ theories arising from $G_2$-holonomy orbifolds.

We begin by noting that, essentially following the discussion of the previous subsection, a {\it generic} element of $\Gamma$ will act on $ {\cal C}_{a_1}$ and  ${\cal C}_{a_2}$ with isolated fixed points because it will act on the base $\mathbb{CP}^1$ fixing two points.
Each of these isolated fixed points itself supports a 5d SCFT. Hence, since $\Gamma$ will contain several such elements, there will be several pairs of such isolated fixed points on the $\mathbb{CP}^1$. Hence, the 5d theory will contain several pairs of 5d SCFTs all coupled together via the BPS instanton particles corresponding to $M$2-branes wrapping the $\mathbb{CP}^1/\Gamma$. Additionally there will also be non-Abelian 5d gauge fields if $\Gamma$ contains elements which fix the $\mathbb{CP}^1$. Finally, if $\Gamma$ also has elements which have codimension four fixed points and act non-trivially on the base $\mathbb{CP}^1$, then there will be {\it pairs} of non-Abelian flavour symmetries present also.

How do these symmetries act on  ${\cal C}_{\mu} = T^*(S^3)$ ? By extending slightly the analysis of section three and using the the fact that the $z_i$ coordinates transform in the fundamental representation of $SO(4)$ one can see that whenever $ {\cal C}_{a_i}/\Gamma$ has a pair of flavour symmetries,  ${\cal C}_{\mu}/\Gamma$ has a codimension four singularity along a single copy of $T^*(S^1) \subset  {\cal C}_{\mu}$. This indicates that the pair of global symmetries in the theories obtained from $ {\cal C}_{a_i}/\Gamma$ get broken to a diagonal subgroup by appropriate vacuum expectation values to give the theory on $ {\cal C}_{\mu}/\Gamma$.

Furthermore, the elements of $\Gamma$ which have isolated fixed points on $ {\cal C}_{a_i}$ tend to act freely on ${\cal C}_{\mu} = T^*(S^3)$. Since they act freely they will produce incontractible cycles which give rise to confining strings. This gives a correspondence between the 5d CFT sectors arising from $M$-theory on ${\cal C}_{a_i}/\Gamma$ and confining strings in $M$-theory on ${\cal C}_{\mu}/\Gamma$.
This also leads to an interesting observation which we will finish with.

Consider the $\mathbb{Z}_p$ action on the conifold coordinates given by:
\be
(X,V,U,Y) \;\; \longrightarrow \;\; (\alpha X, \alpha^q V , \bar{\alpha}^q U , \bar{\alpha} Y)
\ee
where $\alpha^p = 1$ and $q \neq 0 \in \mathbb{Z}$. We will restrict to $p$ and $q$ being relatively prime for simplicity.

Then, ${\cal C}_{\mu}/{\mathbb{Z}_p} = T^*(L(p,q))$, with $L(p,q)$ a general three dimensional Lens space.
In particular, since the first homology group of $L(p,q)= \mathbb{Z}_p$ we have a massive confining theory with strings whose charges are quantised mod $p$. When $|q|=1$ this is exactly the model we obtained by deforming the 5d SCFT which arises from $SU(p)$ super Yang-Mills at strong coupling. 
When $|q| \neq 1$ $M$-theory on $ {\cal C}_{a_i}/\mathbb{Z}_p$ gives a pair of 5d SCFTs coupled to each other via a massive instanton operator. On the other hand 
$M$-theory on $T^*(L(p,q))$ looks like a massive confining theory for all non-zero $p$ and $q$.
It would be interesting to distinguish physically $M$-theory on $T^*(L(p,1))$ from $M$-theory on $T^*(L(p,q))$ with $|q| \neq 1$.

\bigskip
\large
\noindent
{\bf {\sf Acknowledgements.}}
\normalsize

We would like to thank A. Antinucci, D. Baldwin, L. Foscolo, K. Intriligator, J. Lotay and G. Rizi for discussions. In particular, we especially acknowledge discussions with L. di Pietro, whose questions led to the main insight of this paper. The work of BSA is supported by a grant from the Simons Foundation (\#488569, Bobby Acharya)

\bigskip

\bibliographystyle{plain} 

\bibliography{References}

\begin{thebibliography}{10}

\bibitem{bsa2}
Bobby~S Acharya.
\newblock {On realising $N$= 1 super Yang-Mills in $M$-theory}.
\newblock {\em arXiv preprint hep-th/0011089}, 2000.

\bibitem{Acharya:2001hq}
Bobby~S Acharya.
\newblock {Confining strings from G(2) holonomy space-times}.
\newblock {\em arXiv hep-th/0101206}, 1 2001.

\bibitem{Acharya:2023bth}
Bobby~S Acharya, Michele Del~Zotto, Jonathan~J. Heckman, Max Hubner, and Ethan Torres.
\newblock {Junctions, Edge Modes, and $G_2$-Holonomy Orbifolds}.
\newblock {\em arXiv hep-th 2304.03300}, 4 2023.

\bibitem{Acharya:2001dz}
Bobby~S Acharya and Cumrun Vafa.
\newblock {On domain walls of N=1 supersymmetric Yang-Mills in four-dimensions}.
\newblock {\em arXiv hep-th/0103011}, 3 2001.

\bibitem{Aharony:1997bh}
Ofer Aharony, Amihay Hanany, and Barak Kol.
\newblock {Webs of (p,q) five-branes, five-dimensional field theories and grid diagrams}.
\newblock {\em JHEP}, 01:002, 1998.

\bibitem{Apruzzi:2021nmk}
Fabio Apruzzi, Federico Bonetti, I\~naki Garc\'\i{}a~Etxebarria, Saghar~S. Hosseini, and Sakura Schafer-Nameki.
\newblock {Symmetry TFTs from String Theory}.
\newblock {\em Commun. Math. Phys.}, 402(1):895--949, 2023.

\bibitem{Atiyah:2000z}
Michael Atiyah, Juan~Martin Maldacena, and Cumrun Vafa.
\newblock {An M theory flop as a large N duality}.
\newblock {\em J. Math. Phys.}, 42:3209--3220, 2001.

\bibitem{Atiyah:2001qf}
Michael Atiyah and Edward Witten.
\newblock {M theory dynamics on a manifold of G(2) holonomy}.
\newblock {\em Adv. Theor. Math. Phys.}, 6:1--106, 2003.

\bibitem{Candelas:1989js}
Philip Candelas and Xenia~C. de~la Ossa.
\newblock {Comments on Conifolds}.
\newblock {\em Nucl. Phys. B}, 342:246--268, 1990.

\bibitem{Conlon:2022nug}
Ronan~J. Conlon and Hans-Joachim Hein.
\newblock {Classification of asymptotically conical Calabi-Yau manifolds}.
\newblock {\em arXiv math.DG 2201.00870}, 1 2022.

\bibitem{Cordova:2016xhm}
Clay Cordova, Thomas~T. Dumitrescu, and Kenneth Intriligator.
\newblock {Deformations of Superconformal Theories}.
\newblock {\em JHEP}, 11:135, 2016.

\bibitem{Cremonesi:2015lsa}
Stefano Cremonesi, Giulia Ferlito, Amihay Hanany, and Noppadol Mekareeya.
\newblock {Instanton Operators and the Higgs Branch at Infinite Coupling}.
\newblock {\em JHEP}, 04:042, 2017.

\bibitem{Cvetic:2000db}
Mirjam Cvetic, G.~W. Gibbons, Hong Lu, and C.~N. Pope.
\newblock {Ricci flat metrics, harmonic forms and brane resolutions}.
\newblock {\em Commun. Math. Phys.}, 232:457--500, 2003.

\bibitem{Cvetic:2022imb}
Mirjam Cveti\v{c}, Jonathan~J. Heckman, Max H\"ubner, and Ethan Torres.
\newblock {0-form, 1-form, and 2-group symmetries via cutting and gluing of orbifolds}.
\newblock {\em Phys. Rev. D}, 106(10):106003, 2022.

\bibitem{Davies:2011is}
Rhys Davies.
\newblock {Hyperconifold Transitions, Mirror Symmetry, and String Theory}.
\newblock {\em Nucl. Phys. B}, 850:214--231, 2011.

\bibitem{Davies:2013pna}
Rhys Davies.
\newblock {Classification and Properties of Hyperconifold Singularities and Transitions}.
\newblock {\em arXiv hep-th 1309.6778}, 9 2013.

\bibitem{DelZotto:2022fnw}
Michele Del~Zotto, Jonathan~J. Heckman, Shani~Nadir Meynet, Robert Moscrop, and Hao~Y. Zhang.
\newblock {Higher symmetries of 5D orbifold SCFTs}.
\newblock {\em Phys. Rev. D}, 106(4):046010, 2022.

\bibitem{Douglas:1996xp}
Michael~R. Douglas, Sheldon~H. Katz, and Cumrun Vafa.
\newblock {Small instantons, Del Pezzo surfaces and type I-prime theory}.
\newblock {\em Nucl. Phys. B}, 497:155--172, 1997.

\bibitem{Gubser:1998ia}
Steven Gubser, Nikita Nekrasov, and Samson Shatashvili.
\newblock {Generalized conifolds and 4-Dimensional N=1 SuperConformal Field Theory}.
\newblock {\em JHEP}, 05:003, 1999.

\bibitem{Hausel:2002xg}
Tamas Hausel, Eugenie Hunsicker, and Rafe Mazzeo.
\newblock {Hodge cohomology of gravitational instantons}.
\newblock {\em arXiv math/0207169}, 7 2002.

\bibitem{Intriligator:1997pq}
Kenneth~A. Intriligator, David~R. Morrison, and Nathan Seiberg.
\newblock {Five-dimensional supersymmetric gauge theories and degenerations of Calabi-Yau spaces}.
\newblock {\em Nucl. Phys. B}, 497:56--100, 1997.

\bibitem{Lambert:2014jna}
N.~Lambert, C.~Papageorgakis, and M.~Schmidt-Sommerfeld.
\newblock {Instanton Operators in Five-Dimensional Gauge Theories}.
\newblock {\em JHEP}, 03:019, 2015.

\bibitem{Morrison:1996xf}
David~R. Morrison and Nathan Seiberg.
\newblock {Extremal transitions and five-dimensional supersymmetric field theories}.
\newblock {\em Nucl. Phys. B}, 483:229--247, 1997.

\bibitem{Polyakov:1975rs}
Alexander~M. Polyakov.
\newblock {Compact Gauge Fields and the Infrared Catastrophe}.
\newblock {\em Phys. Lett. B}, 59:82--84, 1975.

\bibitem{Lockhart}
ROBERT C. MC~OWEN ROBERT B.~LOCKHART.
\newblock Elliptic differential operators on noncompact manifolds.
\newblock {\em Annali della Scuola Normale Superiore di Pisa, Classe di Scienze 4e série}, 12:409--447, 1985.

\bibitem{Seiberg:1996bd}
Nathan Seiberg.
\newblock {Five-dimensional SUSY field theories, nontrivial fixed points and string dynamics}.
\newblock {\em Phys. Lett. B}, 388:753--760, 1996.

\bibitem{Tachikawa:2015mha}
Yuji Tachikawa.
\newblock {Instanton operators and symmetry enhancement in 5d supersymmetric gauge theories}.
\newblock {\em PTEP}, 2015(4):043B06, 2015.

\end{thebibliography}

\end{document}